\documentclass[modern]{aastex631}

\usepackage{graphicx}
\usepackage{amsmath}
\usepackage{amssymb}
\usepackage{xcolor}
\usepackage{listings}
\usepackage{csquotes}
\usepackage{indentfirst}
\usepackage{comment}
\usepackage{hyperref}
\usepackage{float}

\usepackage{csquotes}
\usepackage{orcidlink}
\usepackage[mathlines]{lineno}

\begin{document}


\title{Spectral Energy Distributions of Globular Clusters in VV191a}

\author{Ashton Cardona \orcidlink{0009-0004-3456-0134} }
\author{Timothy Carleton \orcidlink{0000-0001-6650-2853}}
\author{Jessica Berkheimer \orcidlink{0000-0001-6265-0541}}
\author{Rogier A. Windhorst \orcidlink{0000-0001-8156-6281}}
\author{Seth Cohen \orcidlink{0000-0003-3329-1337}}
\author{Rolf Jansen \orcidlink{0000-0003-1268-5230}}
\affiliation{School of Earth and Space Exploration, Arizona State University, 1151 S Forest Ave Tempe, AZ 85281, USA}

\author{Anton M. Koekemoer \orcidlink{0000-0002-6610-2048}}
\affiliation{Space Telescope Science Institute, 3700 San Martin Drive,
Baltimore, MD 21218, USA}

\author{Gibson Bowling \orcidlink{0009-0007-0782-0721}}
\affiliation{School of Earth and Space Exploration, Arizona State University, 1151 S Forest Ave Tempe, AZ 85281, USA}

\author{Patrick Kamieneski \orcidlink{0000-0001-9394-6732}}
\affiliation{School of Earth and Space Exploration, Arizona State University, 1151 S Forest Ave Tempe, AZ 85281, USA}
\affiliation{Department of Physics and Astronomy, Chalmers University of Technology, SE-412 96 Gothenburg, Sweden}

\author{Jakob Perivolotis \orcidlink{0009-0000-3166-0595}}
\affiliation{Department of Physics and Astronomy, 105 Jesse Hall, Columbia, MO 65211}

\author{Tyler Hinrichs \orcidlink{0009-0008-0376-3771}}
\affiliation{Department of Astronomy, Indiana University, 727 East Third Street, Bloomington, IN 47405, USA}

\author{Christopher Willmer \orcidlink{0000-0001-9262-9997
}}
\affiliation{Steward Observatory, University of Arizona, 933 North Cherry Avenue, Tucson AZ, 85721, USA
}



\begin{abstract}

We investigate globular cluster (GC) candidates associated with the elliptical galaxy VV191a using JWST NIRCam and archival HST imaging. Beginning with an initial catalog of over 500 candidate sources, we apply morphological filtering, photometric redshift selection, and spectral energy distribution (SED) fitting using \texttt{BAGPIPES} to isolate a refined sample of 61 probable GCs, including a high-confidence subset of 20 objects. The derived age distribution is concentrated between $\sim6$--$10$ Gyr, with a peak near $\sim8$ Gyr, while metallicities are predominantly near and super-solar. Unlike many GC systems in elliptical galaxies, the VV191a sample does not exhibit a clear bimodal metallicity distribution. Derived stellar masses span a broad range and suggest significant degeneracies related to dust attenuation and foreground contamination from the overlapping spiral galaxy VV191b. These results support the presence of an evolved GC population in VV191a while highlighting limitations imposed by broadband SED fitting and foreground contamination.
\end{abstract}

\section{Introduction}
Globular clusters (GCs) exist as some of the oldest stellar systems in galaxies and serve as powerful tracers of their evolution across cosmic time. Derived cluster properties encode information about conditions of their formation, making them useful probes of a galaxy's assembly history \citep{Kruij2015, 2020beasley}. GCs in elliptical galaxies often exhibit characteristic features that reflect multiple epochs of star formation and growth via mergers.

 Evidence has been found in previous work of probable GCs in VV191a \citep{Berkheimer_2024}. In this work, we refine a catalog of candidate objects and model their spectral energy distributions (SEDs) to constrain key properties such as age, metallicity, and stellar mass. By analyzing these properties, we aim to assess the nature of the GC population in VV191a.



\section{Data and Methods}


The VV191 system presents a unique opportunity to study such a population in a more complex observational environment, enabling investigation of GCs in the elliptical while simultaneously introducing observational challenges due to contamination from the foreground spiral galaxy VV191b. We utilize imaging of the VV191 system (at DOI: \dataset[10.17909/9a9t-3q60]{https://doi.org/10.17909/9a9t-3q60}) obtained with the \textit{James Webb Space Telescope} (JWST) NIRCam instrument in the F090W, F150W, F356W, and F444W filters as part of the PEARLS survey found in \citet{Windhorst2023} (PID: 1176). This data is complemented by archival \textit{Hubble Space Telescope} (HST) imaging in the F606W filter taken from MAST (PID: 13695). The JWST data was processed using the standard calibration pipeline \citep{Bushouse2023} and combined into mosaics with a pixel scale of $0.03''$.

We began with a catalog of $\sim$500 sources identified in \citet{Berkheimer_2024}, which includes candidate GCs, foreground, and background objects. Due to observed substantial overlap of the spiral arms associated with VV191b, it is expected that there will be significant contamination from foreground and background objects \citep{Keel2023}. As such, a multi-stage procedure was implemented to isolate and select a robust candidate GC sample.

We performed aperture photometry for each candidate across all available filters using the \texttt{PhotUtils} aperture photometry routines. Background subtraction was performed using annuli with inner and outer radii of $0.5\arcsec$ and $0.75\arcsec$, respectively. Uncertainties were similarly calculated directly. We compared resulting SEDs to model predictions for compact star clusters from \citet{Faisst2022}. Sources displaying irregular photometric behavior inconsistent with GC expectations were excluded from the final sample.

Following morphological filtering, we computed photometric redshifts for the remaining candidates using the \texttt{EAZY-Py} package \citep{eazypy}. Each object is compared against the known redshift of VV191a ($z \approx 0.0513$), with a relatively broad tolerance threshold of $\Delta z = \pm 0.15$ to account for broadband degeneracies. 

As an additional consistency check, we examine the spatial distribution of selected sources via their placement on a color--magnitude diagram, where genuine GCs are expected to occupy a relatively narrow locus reflecting evolved stellar populations.

We construct and model the SED of each remaining candidate using the \texttt{BAGPIPES} fitting framework \citep{bagpipes}. This method fits stellar population synthesis models to the observed photometry, allowing us to infer key physical parameters such as stellar age, mass, and metallicity.

The fitting procedure explores a wide parameter space in age (1-14 Gyr), metallicity ($\log Z/Z_{\odot} =$ 0 - 2), mass ($10^3$-$10^{15}$ $M_{\odot}$), and dust attenuation ($0<A_V<1$), with uniform priors, to avoid biasing toward specific solutions.

The quality of each SED fit is evaluated using a $\chi^2_{phot}$ statistic, and objects with $\chi^2_{phot} > 50$ are excluded as poorly constrained or likely contaminants. Even for well-fit objects, significant degeneracies exist between parameters such as age, metallicity, and dust attenuation \citep{conroy2013, worthey1994, papovich2001}. After performing this check of the $\chi^2$, we obtain a high-confidence sample of 20 candidate objects.

\section{Results}

\begin{figure}[H]

    \includegraphics[width=0.9\textwidth]{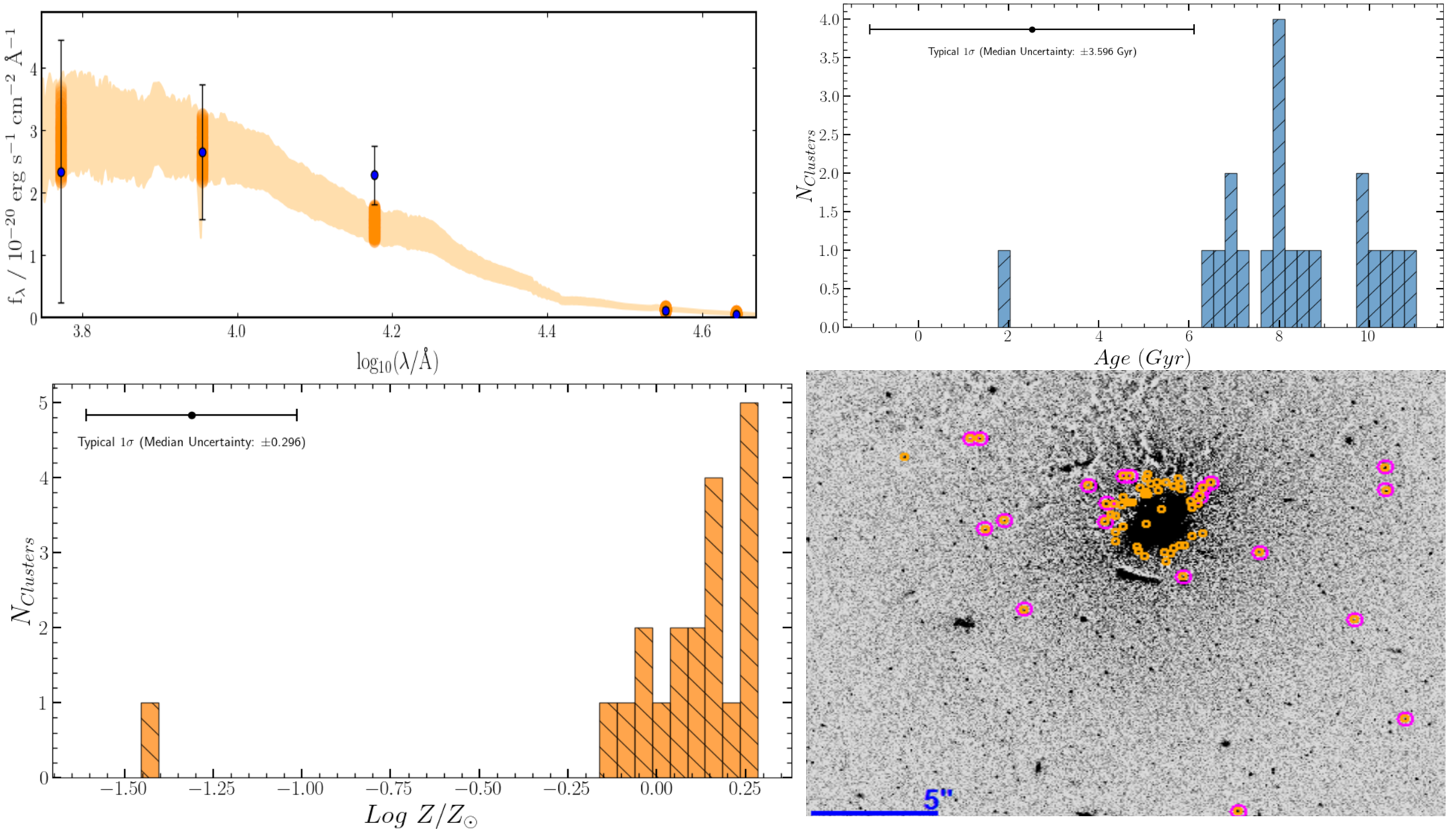}
    \caption{Top Left: Representative SED fit for a high-confidence GC candidate in VV191a. The declining spectral shape toward longer wavelengths is consistent with an evolved stellar population. Top right: Distribution of derived cluster ages for the high-confidence sample, showing a concentration between $\sim6$--$10$ Gyr with a peak near $\sim8$ Gyr. Bottom left: Metallicity distribution of the same sample, dominated by near- and super-solar metallicities with only a small number of lower-metallicity outliers. Bottom right: DS9 image of background-subtracted VV191a with initial and high-confidence samples' spatial distributions, orange circles represent the overall sample of 61 candidates, while magenta are the 20 high-confidence objects.}
    \label{fig:distributions}
\end{figure}

SED Fitting of the high-confidence candidates from \texttt{BAGPIPES} in general reproduced observed photometry well across available wavelengths. An example of the fitted SEDs is shown in Figure~\ref{fig:distributions}, with the best-fit model capturing declining spectral shapes anticipated for an evolved stellar population dominated by older stars.

Figure~\ref{fig:distributions} additionally shows derived ages of the 20 high-confidence objects. The age distribution centers on a concentration at $\sim 6-10$ Gyr, peaking at 8 Gyr, which is slightly younger than the canonical $\gtrsim10$ Gyr. 

The metallicity distribution is strongly skewed toward peaks at near-solar and super-solar metallicity, with a very small number trending toward lower metallicity values. Unlike a majority of GC systems in elliptical galaxies, this sample from VV191a does not show an expected bimodal distribution \citep{Forbes1997, Harris1999}. The age--metallicity relation generally predicts metal-rich GCs form early and are rapidly enriched \citep{leaman2013}. However, intermediate-aged ($\sim$8 Gyr), metal-rich GCs have been identified in merger-remnant and lenticular galaxies \citep{puzia2003}; this suggests VV191a may have experienced a more extended GC formation history. Alternatively, age--metallicity degeneracies might bias estimates toward younger, metal-rich solutions. 

Masses derived from the fits suggest these objects span a range between $\log(M/M_\odot)$ $\sim$ 5 to 9.5. There is a slight peak at $\log(M/M_\odot)$ $\sim$ 7 – 7.5, but the distribution is very wide with no obvious Gaussian or bimodal shape. Results from this high-confidence sample can be found at \url {https://tinyurl.com/VV191GCs}

\section{Conclusion}

We refined a sample of GC candidates associated with the galaxy VV191a using JWST and HST photometry, using morphological filtering, photometric redshift checks, and SED modeling. We investigated the properties of the refined sample of 61 candidates, with a high-confidence sample of 20 probable GCs.

Despite limitations due to broadband photometry and parameter degenracies, the results support the presence of a substantially evolved GC population associated with VV191a. Future spectroscopic observations and deeper multi-wavelength imaging will be essential for confirming cluster membership and improving constraints on the formation history and structure of the VV191a GC system.

\section{Acknowledgments}
This research is based on observations made with the NASA/ESA Hubble Space Telescope obtained from the Space Telescope Science Institute, which is operated by the Association of Universities for Research in Astronomy, Inc., under NASA contract NAS 5–26555. These observations are associated with program 13695.

\newpage
\bibliography{bib}

\end{document}